# DIRC-like time-of-flight detector for the experiment at the Super Tau-Charm Facility

**Binbin Qi[a,b], Ziwei Li[a,b], Ming Shao[a,b,1], Jianbei Liu[a,b,1], Zhujun Fang[a,b], Huangchao Shi[a,b] and Xin Li[a,b]**

[a] *State Key Laboratory of Particle Detection and Electronics, University of Science and Technology of China, Hefei 230026, China*

[b] *Department of Modern Physics, University of Science and Technology of China, Hefei 230026, China*
  *E-mail:* swing@ustc.edu.cn (Ming Shao); liujianb@ustc.edu.cn (Jianbei Liu)

ABSTRACT: The Super Tau-Charm Facility (STCF) is a future electron-positron collider proposed in China with a peak luminosity of above $0.5 \times 10^{35}$ cm$^{-2}$s$^{-1}$ and center-of-mass energy ranging from 2 to 7 GeV. An excellent particle identification (PID) capability is one of the most important requirements for the detector at the STCF. A 3σ π/K separation power at the momentum of up to 2 GeV/c is required within the detector acceptance. A DIRC-like time-of-flight (DTOF) detector is proposed to meet the PID requirement for the endcap region of the STCF. The conceptual design of the DTOF detector and its geometry optimization is herein presented. The PID performance of the detector is studied using Geant4 simulation. With a proper reconstruction algorithm, an overall time resolution of ~50 ps is achieved for the detector with an optimum geometry when convoluting contributions from all other sources, including the transit time spread (TTS) of the photodetector, electronic timing accuracy, and an assumed precision (~40 ps) of the event start time. A π/K separation power of better than 4σ at the momentum of 2 GeV/c is achieved over the entire sensitive area of the DTOF detector, thereby fulfilling the physics requirement of the PID detector for the experiment at the STCF.

---

[1] Corresponding authors.

# Contents

## 1. Introduction

The Super Tau-Charm Facility (STCF, [1,2]) is a future high-luminosity electron-positron collider that was proposed in China. This collider is expected to operate in a center-of-mass energy region ranging from 2 to 7 GeV with a peak luminosity of $>0.5 \times 10^{35}$ cm$^{-2}$s$^{-1}$. The energy region covered by the STCF lies in the transition interval between non-perturbative quantum chromodynamics (QCD) and perturbative QCD, enabling a rich physics program, including τ and charm physics, hadron physics, and new physics searches.

Excellent particle identification (PID) capability is a crucial demand in terms of detector performance driven by the physics program at the STCF. According to physics case studies for the STCF, a statistical separation power better than 3σ between charged hadrons ($\pi^\pm$, $K^\pm$, and $p/\bar{p}$) at the momentum of up to 2 GeV/c is required within the entire detector acceptance [2]. In the physics research conducted at STCF, charged hadrons at a low momentum can be identified with the characteristic ionization energy loss (dE/dx) using a tracking detector. However, at a higher momentum, a dedicated PID detector is needed to accomplish the required particle separation capability. The PID detector must also be able to cope with the high radiation level and counting rate anticipated in the high luminosity condition at the STCF. Therefore, it should have a fast response and good radiation resistance. Furthermore, the material budget of a PID detector should be kept as low as possible to minimize its impact on the energy measurement of photons with an electromagnetic calorimeter (EMC) outside the PID detector. Ultimately, the PID detector should be sufficiently compact to be accommodated in the rather limited space between the tracking detector and the EMC.

Because of the limited distance between the interaction point and the barrel of the PID detector at the STCF, the usual time of flight (TOF) method cannot be applied to the barrel region given the required PID capability. The Ring Imaging Cherenkov detector (RICH, [3]) is therefore a good candidate for the barrel PID detector. However, a TOF detector with very high intrinsic time resolution and fast response is a viable option for the endcap PID detector, owing to the extended distance between the interaction point and the endcap PID detectors. The concept of Detection of Internal total-Reflected Cherenkov light (DIRC, [4]) offers a solution in this regard. The DIRC-like time of flight detector (DTOF) was proposed as the endcap PID detector in the STCF experiment towing to its compact structure, simplified operation and maintenance, high rate capability and high radiation tolerance, and excellent timing performance.

## 2. DTOF detector concept

The concept of DIRC was first introduced in the Babar experiment [5,6]. In this concept, Cherenkov photons generated in a long-fused silica bar were propagated to one end through total internal reflections and then projected to an array of photo sensors via a water expansion volume. The fused silica bar was used as both the Cherenkov radiator and the light guide. The propagation direction of the Cherenkov photon was preserved through hundreds of reflections, and the spatial pattern of the Cherenkov ring was recognized for PID purpose. Notably, the time resolution of the Babar DIRC detector for a single photon was approximately 1 ns, and this was mainly used to suppress uncorrelated backgrounds by setting a proper time window. Consequently, the DIRC detectors with very fast timing capability were developed to explore the PID potential of detectors in the time domain. One example in this regard is the imaging Time of Propagation (iTOP) detector that was used in the Belle II [7-9] experiment, where the PID performance was primarily driven by a precise measurement of the propagation time of Cherenkov photons transmitted in the cuboid-shaped radiator plate. As a result, the two-dimensional position and the one-dimensional time of the photon hits were measured using an array of photo sensors at one end of the plate, with a particularly high precision timing resolution. Some compact DIRC detectors can be realized by improving the position and time resolution and are expected to have better PID capability, such as those proposed in the future PANDA experiment [10-12] and EIC project [13,14]. The excellent timing capability of the new DIRC detectors has also led to the direct application of the DIRC technology to high time resolution and high rate TOF measurement. Recent examples of such DIRC-like detectors include the FTOF detector proposed for the superB project [15,16], the TORCH detector that is being developed for LHCb upgrade [17,18], and the DTOF detector that was proposed for the STCF experiment.

The DTOF detector requires a total time resolution of 50 ps in the TOF measurement in order to provide a $3\sigma$ $\pi/K$ separation power at the momentum of up to 2 GeV/c for the STCF experiment. It is located in the endcap region of the STCF detector and comprises two identical discs positioned at 1400 mm along the beam direction away from the collision point. Each disc is made up of a few sectors (shown in Fig. 1), with an inner radius of ∼560 mm and outer radius of ∼1050 mm, covering ∼22°-36° in terms of the polar angle θ. In each sector, a synthetic fused silica plate is used as the radiator to generate Cherenkov photons. Considering the effect of magnetic field on the photon sensors, an array of multi-anode micro-channel plate photomultipliers (MCP-PMT) are optically coupled to the radiator along the outer side. The entire sector is enclosed in a light-tight black box, occupying ∼200 mm space along the beam direction. The other parameters, including the number of sectors, thickness of the plate, the number of sensors, and different mirror settings will be described and studies in section 4.

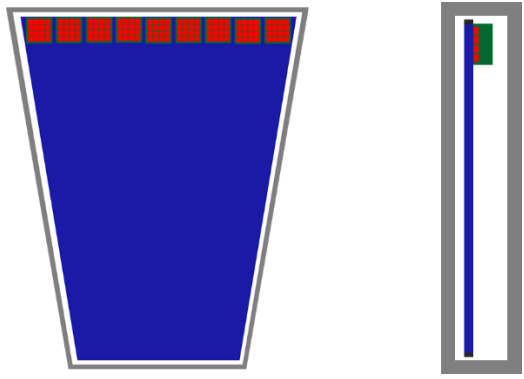

**Figure 1.** Conceptual design of the DTOF. The front (left) and side (right) views of a sector are shown. A fused silica plate is used as a radiator, and an array of MCP-PMTs, coupled to the radiator, is used as the photon sensor. The entire sector is enclosed in a light-tight black box.

## 3. DTOF timing uncertainty analysis

Time resolution is a key indicator of the performance of a DTOF detector. It is necessary to analyze the factors that affect the timing uncertainty, and the relative importance of the various factors must be investigated to optimize the time performance. Similar to the FTOF detector that was proposed for the superB project, the main sources contributing to the timing uncertainty of the DTOF detector are [15]

$$\sigma_{tot}^2 \sim \sigma_{trk}^2 + \sigma_{T_0}^2 + \left(\frac{\sigma_{elec}}{\sqrt{N_{p.e.}}}\right)^2 + \left(\frac{\sigma_{TTS}}{\sqrt{N_{p.e.}}}\right)^2 + \left(\frac{\sigma_{det}}{\sqrt{N_{p.e.}}}\right)^2,$$

where $N_{p.e.}$ is the number of photo-electrons (p.e.), $\sigma_{trk}$ is the error caused due to track reconstruction, $\sigma_{T_0}$ is the event reference time ($T_0$ is when physical collision happens) error that is mainly affected by the collider design of the STCF, $\sigma_{elec}$ is the electronic timing accuracy, $\sigma_{TTS}$ is the single-photon transit time spread (TTS) of the MCP-PMT, and $\sigma_{det}$ is the time reconstruction uncertainty of the DTOF detector. From this formula, it can be observed that the contribution from $\sigma_{elec}$, $\sigma_{TTS}$, and $\sigma_{det}$ decreases with increasing $N_{p.e.}$, whereas the timing errors from $\sigma_{trk}$ and $\sigma_{T_0}$ remain constant. It will also be noted that the uncertainty of $T_0$ is usually approximately 30~40 ps[1], rendering it an important timing error source at STCF.

To estimate the intrinsic timing uncertainty of DTOF, i.e.

$$\sigma_{DTOF} = \frac{\sigma_{det} \oplus \sigma_{TTS} \oplus \sigma_{elec}}{\sqrt{N_{p.e.}}},$$

we study the main contributing factors by considering a simple case. If a relativistic charged particle is vertically incident on a thin Cherenkov radiator plate, then a Cherenkov light cone will be produced. Assuming the particle moves along the z-direction, the incident point is at the origin (0,0,0), and the photon sensor is located at (X, Y, 0), the distance from the incident point to the photon sensor will be $D = \sqrt{X^2 + Y^2} = L \sin \theta_c$, where L is the length of propagation (LOP) of the Cherenkov photon, and $\theta_c$ is the Cherenkov radiation angle. The time of propagation (TOP) of a photon is given by

$$TOP = \frac{L n_g}{c} = \frac{D n_g}{c \sin \theta_c} = \frac{n_p n_g \beta D}{c\sqrt{n_p^2 \beta^2 - 1}} = \frac{n_p n_g p D}{c\sqrt{n_p^2 p^2 - p^2 - m^2}},$$

---

[1] A typical value for the T0 determination at the BESIII experiment.

where $n_p$ and $n_g$, respectively, are the phase and group refractive indexes of fused silica, which in turn are related to the wavelength λ, c is the velocity of light in vacuum, and $\beta$ and p are the reduced velocity and momentum of the incident particle, respectively. The excitation time of Cherenkov radiation can be deduced using $T_{det} = T - TOP$, where T is the time of photon hits measured at the photon sensor.

By differentiating the abovementioned equation, the sources of the timing error emerge as

$$\sigma_{DTOF}^2 = \sigma_T^2 + \left(\frac{TOP}{D}\right)^2 \sigma_D^2 + \left[\frac{TOP}{n_p n_g}\frac{d(n_p n_g)}{d\lambda} - \frac{TOP^3 c^2}{n_p n_g^2 D^2}\frac{dn_p}{d\lambda}\right]^2 \sigma_\lambda^2 + \left[\frac{TOP}{p} - \frac{TOP^3 c^2 (n_p^2-1)}{n_p^2 n_g^2 p D^2}\right]^2 \sigma_p^2,$$

where $\sigma_T$ is the single photo-electron (SPE) time resolution of the photon sensor and the electronics (also called $\sigma_{SPE}$), $\sigma_\lambda$ is the chromatic dispersion effect, $\sigma_D$ is the position resolution owing to finite photon sensor size, and $\sigma_p$ is the measurement error of particle momentum. To quantitatively estimate $\sigma_{DTOF}$, we assume a sensitive wavelength range of 300-650 nm for photon sensor. The dispersion effect in this wavelength range is $\sigma_\lambda$~75 nm with a mean wavelength of 405 nm. The photon sensor along with the readout electronics produce a timing uncertainty of $\sigma_{SPE}$=70 ps. The finite photon sensor pixel size of 5.5 mm gives a position error of $\sigma_D = 1.6 mm$. The momentum error contribution is considered by a kaon of p = 1 GeV/c. Furthermore, the thickness of fused silica plate (15 mm) also contributes an uncertainty to the excitation time of Cherenkov radiation, estimated as $\frac{Thick}{\beta c \sqrt{12}}$. The multiple Coulomb scattering (MCS) effect causes an angular resolution, which is significant at low momentum (2.8 mrad for a kaon at 1GeV/c). Its impact to time uncertainty is estimated as $\frac{TOP}{\tan\theta_c}\sigma_{MCS}$.

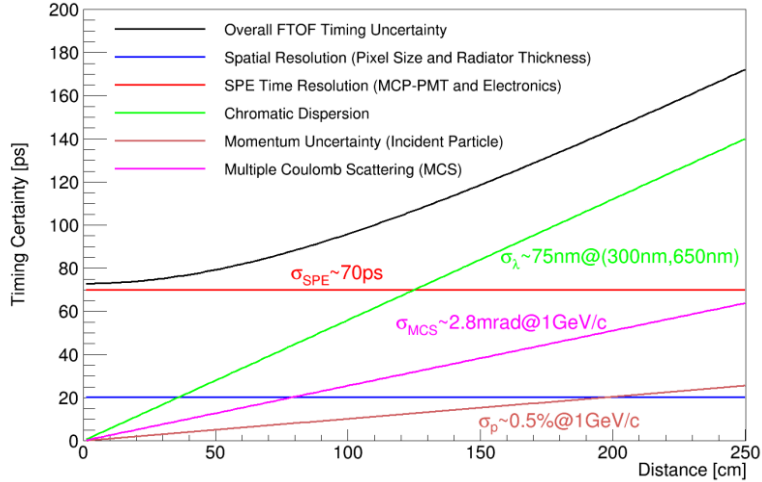

**Figure 2.** Main DTOF timing error factors and their dependences on the distance from the particle (kaon @ p=1GeV/c) incident point to the photon detector.

The calculation results for the SPE are shown in Fig. 2, as a function of the photon transmission distance D. It can be seen that the timing jitter of the photon sensor plays a major role when D is relatively short (< 1 m), whereas the dispersion effect gradually becomes the dominant factor when D is large (> 1.5 m). A proper optical design can be used; then the dispersion effect can be corrected by position information if a very precise timing is maintained. An example is the TORCH with a focusing component [17,18]. For the DTOF detector, the typical D value is approximately 0.5-1 m; hence, the timing uncertainty due to the dispersion effect is smaller than the timing jitter of the photon sensor. This means that the compact design of the DTOF with no focusing component is desirable. In addition, it can be seen that the spatial resolution, including the thickness of fused silica and the pixel size of the photon sensor, has little effect on the time uncertainty of the DTOF. Therefore, a large pixel size photon sensor can be used to reduce the number of electronics readout channel. Notably, the p.e. number may also increase with the thickness of the radiator; however, this can cause an increase in the material budget, although it has little influence on the time resolution of SPE.

## 4. DTOF geometry optimization

Geometry optimization of the DTOF detector has been studied using Geant4 [19] simulation (see section 5.1 for details). The timing performance of different geometry configurations obtained using a reconstruction algorithm (see section 5.2) and their π/K separation power has also been investigated by a likelihood method (see section 5.3). Some geometry configurations are listed in Table 1. We compare the DTOF performances with these different geometry configurations and study the effects derived from three main factors: radiator shape/size, radiator thickness, and the setting of mirrors.

**Table 1.** Description of the different DTOF geometry configurations, where A stands for Absorber and M for Mirror.

| Configuration/Geometry ID | 0 | 1 | 2 | 3 | 4 | 5 | 6 |
|---|---|---|---|---|---|---|---|
| Radiator shapes (sector number) | 4 | 12 | 24 | 4 | 4 | 4 | 4 |
| Radiator thickness (mm) | 15 | 15 | 15 | 10 | 20 | 15 | 15 |
| Outer side surface | A | A | A | A | A | M | 45°M |
| Inner side surface | A | A | A | A | A | A | A |
| Lateral side surface | M | M | M | M | M | M | M |

### 4.1 Configurations

As shown in Fig. 3, we studied three different radiator shapes (and sizes), and they are presented in Geometry 0, 1, and 2. The DTOF disc of Geometry 0 comprises 4 quadrant sectors, in which the planar radiator is fan-shaped and can be viewed as a composite structure of 3 trapezoidal units. Each unit is ~295 mm (inner side) /~533 mm (outer side) wide, ~470 mm high, and 15 mm thick. An array of 3×18 MCP-PMTs are optically coupled to the radiator along the outer side. The size of the MCP-PMT is referred to R10754-07-M16 [20], which is a mature MCP-PMT produced by Hamamatsu. One MCP-PMT contains 4×4 anodes, each with a size of 5.5×5.5 mm$^2$. The disc of Geometry 1 comprises 12 trapezoidal sectors, each with an area that is approximately one-third of Geometry 0. Each sector contains 18 MCP-PMTs. For Geometry 2, the disc contains 24 trapezoidal sectors, each coupled with 8 MCP-PMTs. For convenience, the readout channels of these MCP-PMTs are numbered along the azimuth and radial axes, called X and Y channel ID, respectively. Thus, each sector of Geometry 0, 1 and 2 contains 216×4, 72×4 and 32×4 channels (X×Y), respectively.

The effect of the radiator thickness is analyzed by comparing Geometry 0, 3 and 4. The radiator thicknesses are 15 mm, 10 mm and 20 mm, respectively. For all the geometry configurations listed in Table 1, the inner- and lateral-side surfaces of the radiators are covered with an absorber or a reflective mirror, named as "A" and "M", respectively. On the outer-side surface, a mirror is used to extend the acceptance of the Cherenkov light, thereby increasing the detected number of photons. As shown in Fig. 4, Geometry 5 has a mirror on the outer-side surface of its radiator, and this is equivalent to putting a mirror MCP-PMT parallel to the real one. Geometry 6 has a mirror on the 45°chamfer, which is equivalent to putting a mirror MCP-PMT perpendicular to the real one. Notably, the addition of mirror MCP-PMT will lead to an increase in the number of possible light paths, which may affect the reconstruction, and ultimately, the time resolution.

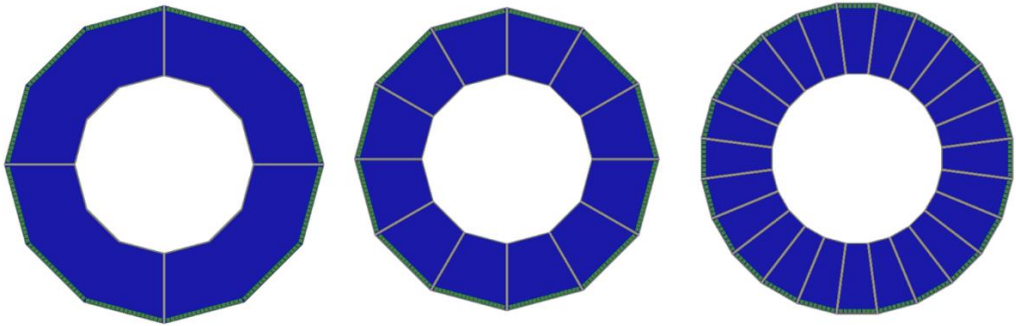

**Figure 3.** Three different radiator shapes. The disc of these configurations contains 4 (left), 12 (center) and 24 (right) sectors, respectively.

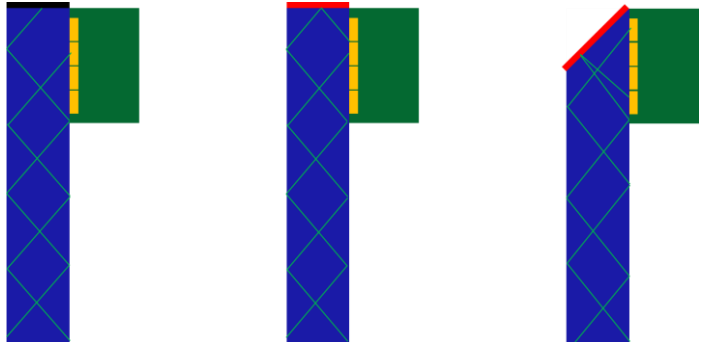

**Figure 4.** Three different configurations on the outer surface of the radiator. An absorber (left) or mirror (center) on the outer surface, and a mirror on the 45°chamfer of the outer side surface (right).

### 4.2 Comparison

The key results of the geometry optimization are listed in Table 2. The number of photoelectrons is obtained from the Geant4 simulation (see section 5.1). For most geometry configurations, except Geometry 5 and 6, a better π/K separation power can be obtained using more photoelectrons. The accumulated charge density on the MCP-PMT anode is estimated from the background study (see section 6 for details), and the π/K separation power is estimated using a likelihood method (see section 5.3).

The effect of the radiator shape/size is investigated by comparing the DTOF performance with Geometry 0, 1 and 2. The p.e. yields of Geometry 0 and 1 are similar and significantly larger than that of Geometry 2. This is because Geometry 2 has less MCP-PMTs per unit radiator area owing to a greater dead area between the sectors. Furthermore, a small radiator increases the light reflection times off the lateral-side mirror, leading to more photon losses. As shown in Fig. 5, the different branches on the time-position hit pattern represent the light propagation from different paths, i.e., photons directly hit the MCP-PMT or reflect off the lateral-side mirror once, twice, or more. In our simulation, the maximum number of reflections from the lateral mirror of Geometry 0, 1, and 2 are one, two and four, respectively. Such reflections cause an overlap on the different hit branches and increase the timing uncertainty. Geometry 0 has the best π/K separation power of 4.17σ, whereas Geometry 2 has the worst at 3.66σ. In addition, a large radiator can significantly reduce dead area, indicating that Geometry 0 is a better choice.

**Table 2.** Performance of different geometries at p=2 GeV/c, θ = 24° and φ = 45°.

| Performance/Geometry ID | 0 | 1 | 2 | 3 | 4 | 5 | 6 |
|---|---|---|---|---|---|---|---|
| Number of photoelectron by pions (except background) | 21.8 | 21.9 | 17.0 | 15.5 | 25.7 | 33.2 | 38.7 |
| Accumulated charge density on MCP-PMT anode (C/cm$^2$) | 10.8 | 10.5 | 9.6 | 8.8 | 11.8 | 17.0 | 25.6 |
| π/K separation power | 4.17σ | 4.08σ | 3.66σ | 3.99σ | 4.27σ | 4.26σ | 4.19σ |

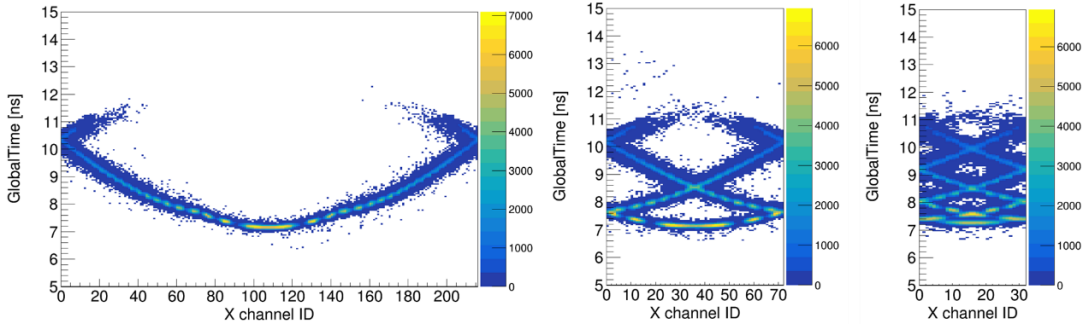

**Figure 5.** Time of the photoelectron arrival vs X channel ID by pions at p=2 GeV/c, θ = 24°, φ = 45°, for different geometries: geometry 0 (left), geometry 1 (center) and geometry 2 (right).

Geometry 3 and 4 have different radiator thicknesses compared to Geometry 0. Although the number of photons generated in the radiator is generally proportional to its thickness, the p.e. yields of Geometry 3 and 4, which are ~16 and ~26 respectively, are not proportional to their radiator thicknesses. This is because the sizes of MCP-PMTs of these geometries are the same. It implies that a thicker radiator increases the probability of light absorption by the outer-side absorber, which in turn indicates a smaller photon acceptance. From the calculation results presented in section 3, we know that the thickness of the radiator has an insignificant effect on the time uncertainty of the SPE. Therefore, more detected photons result in better time resolution. Based on our evaluation, the 15 mm-thick radiator proved to be the most effective owing to its ability to minimize the impact of DTOF material budget on EMC while providing better performance to reduce detector aging in long-term operation.

To increase the p.e. yield, a mirror can be attached to the outer-side surface of the radiator in different ways, as shown in Geometry 5 and 6. The number of p.e. obtained from the two geometries are ~33 and ~39, respectively, and their numbers are higher than that in the case of Geometry 0. However, the outer-side mirror increases the number of possible light paths, which creates a "confusion" similar to the effect of multiple reflections off the lateral-side mirror and further degrades the time resolution. Therefore, the π/K separation powers of Geometry 5 and 6 are similar to Geometry 0, even with more detected photons. In addition, the accumulated charge densities of these two geometry configurations are much higher, adversely affecting the lifetime of MCP-PMT [21,22]. As a result, we rejected the options, where mirrors are attached to the outer-side surface of the radiator, and an optimum geometry configuration of DTOF, i.e. Geometry 0, was obtained and chosen as our baseline design. The results presented in this study were obtained from Geometry 0.

## 5. DTOF PID Performance

### 5.1 Geant4 Simulation

Geant4 [19] simulations are performed to study the expected performance of the DTOF. A 20 mm thick Aluminum plate is added 100 mm in front of the DTOF detector to simulate the material budget of the tracker endcap. Each DTOF sector is enclosed in a light-tight black box composed of 5 mm thick carbon fiber and occupying ~200 mm space along the direction of the beam tube. When tracking the Cherenkov photon propagation, the two lateral sides of radiator are set reflective with a reflection coefficient of ~92%. The surface roughness of the radiator is simulated by randomizing the normal direction of the facet by $\sigma_\alpha=0.1°$, and set with a reflection coefficient of ~97%. The windows of the photon detector are directly coupled to the surface of the fused silica with no air gap or other cookies, and a quantum efficiency is provided to simulate the response of the photon detector (refer to Hamamatsu R10754-07-M16 [20]). All optical parameters used in simulation, including refraction index, reflection coefficient of mirror and radiator surface, and photon absorption length in radiator are wavelength-dependent to simulate the chromatic dispersion. Pions and kaons are emitted from the interaction point at different momenta and directions. A typical Cherenkov photon hit pattern is generated as pions at p=2 GeV/c, $\theta = 24°$ and $\varphi = 15°$, as displayed in Fig. 6. A clear correlation between time and the hit position (sensor pixel) is demonstrated. Two bands, representing photons reflect off the lateral mirror zeroth and once, respectively, are well separated except for a few sensors close to the lateral mirror. The mean number of photons detected by the MCP-PMT arrays is ~21, also shown in Fig. 6. To further study the time resolution of the DTOF detector, a reconstruction algorithm is required, as shown below.

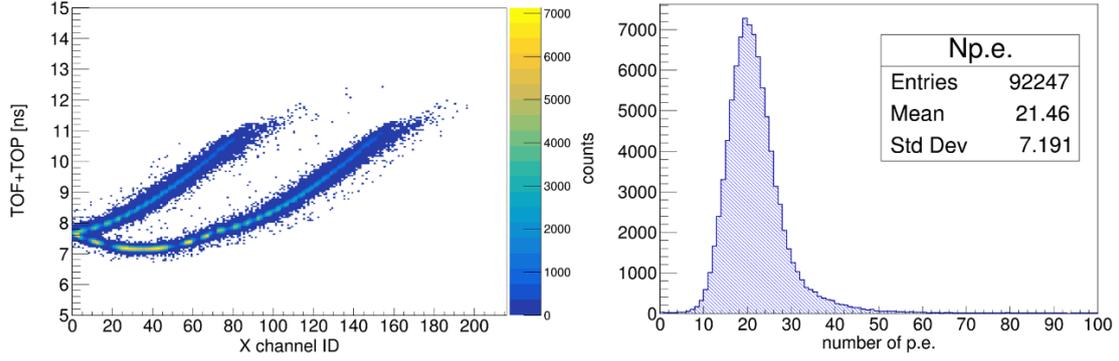

**Figure 6.** Simulated TOP vs. hit position pattern of DTOF, and the number of photoelectrons detected by the MCP-PMT array.

### 5.2 Reconstruction algorithm

The DTOF reconstruction is performed using the coordinate system for one DTOF quadrant, as shown in Fig. 7. According to the Cherenkov angle relation $cos(\overline{\theta}_c) = \frac{1}{n_p \beta} = \frac{\vec{v_t} \cdot \vec{v_p}}{|\vec{v_t}| \cdot |\vec{v_p}|}$, $\vec{v_t} = (a, b, c)$ is the velocity vector of the incident particle when impinging the radiator, $\vec{v_p}$ is the velocity vector of the emitted Cherenkov photon, $n_p$ is the refractive index of the radiator, and $\beta$ is the reduced speed of the particle (set a hypothetical particle to calculate $\beta$). The directional components of $\vec{v_p}$ can be expressed as $(\Delta X, \Delta Y, \Delta Z)$, representing the 3D spatial difference between the photon sensor pixel and the incident position on the radiator surface, as depicted in Fig. 7 (right). Although the 2D (X and Y) difference can be readily obtained, $\Delta Z$ must be deduced using a particle-species hypothesis. If $V = cos(\overline{\theta}_c)$ is known, the equation regarding $\Delta Z$ can be expressed as follows:

$$(c^2 - V^2)\Delta Z^2 + 2c(a\Delta X + b\Delta Y)\Delta Z + (a\Delta X + b\Delta Y)^2 - V^2(\Delta X^2 + \Delta Y^2) = 0$$

By solving this equation, we find $\Delta Z = \frac{-B \pm \sqrt{B^2 - 4AC}}{2A}$, with $A = c^2 - V^2$, $B = 2c(a\Delta X + b\Delta Y)$ and $C = (a\Delta X + b\Delta Y)^2 - V^2(\Delta X^2 + \Delta Y^2)$. In order to obtain a real solution, $\Delta = B^2 - 4AC \geq 0$ is required. After some further physical cuts, $V > 0$ (Cherenkov photon forwardly emitted) and $\frac{\Delta x^2 + \Delta y^2}{\Delta x^2 + \Delta y^2 + \Delta z^2} \geq \frac{1}{n_p^2}$ (ensuring internal total reflection) are applied, the minimal solution ($\Delta z = \min(|\Delta z_1|, |\Delta z_2|)$) is taken as the optimum solution.

The LOP of photons inside the radiator is obtained using the equation $LOP = \sqrt{\Delta X^2 + \Delta Y^2 + \Delta Z^2}$. The precision of the reconstructed LOP is ~3.3 mm, by pions at p = 2 GeV/c, θ = 24° and φ = 45°, as shown in Fig. 8. Furthermore, we find that the reconstruction algorithm is well suited for most photon sensors regardless of the incident position of the particles.

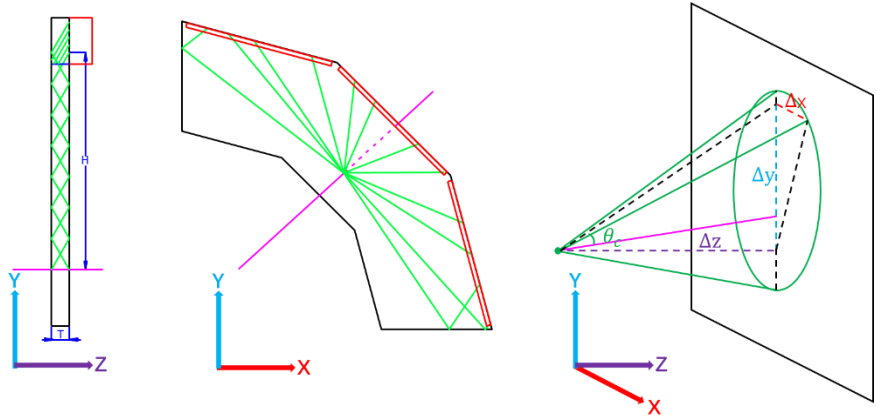

**Figure 7.** Coordinate system used in DTOF reconstruction (left and center) and the direction of the Cherenkov photon (right), the green line represents the Cherenkov photon whereas the pink line represents the incident charged particle.

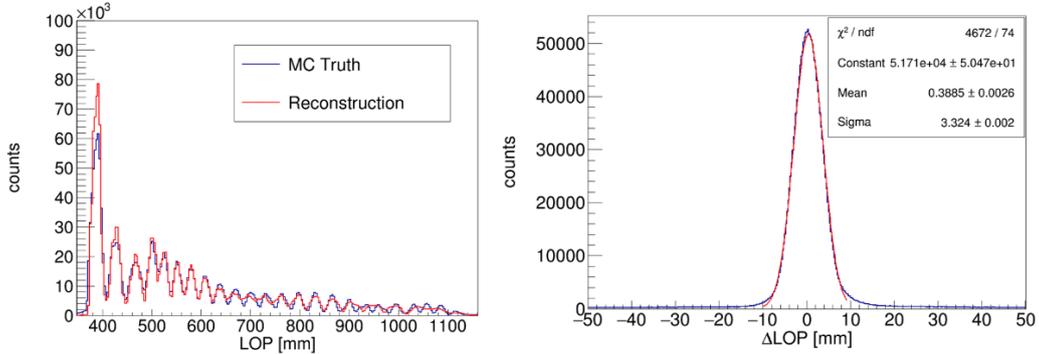

**Figure 8.** Reconstructed propagation length of Cherenkov photons (left) and its uncertainty (right) in the fused silica plate of DTOF.

The TOF information is obtained applying the following formula: $TOF = T - TOP - T_0 = T - \frac{L\overline{n_g}}{c} - T_0$. As shown above, the light may take different paths to the same pixel sensor; hence, the TOFs of all possible paths are reconstructed, and the one closest to the hypothetical value is retained. Fig. 9 shows the time resolution of the DTOF detector by pions at p = 2 GeV/c, θ = 24° and φ = 45°, for the SPE and combining all photoelectrons, respectively. The timing jitter of the MCP-PMT and electronics are not considered. For the SPE, the intrinsic time resolution from DTOF reconstruction is ~41 ps. Averaging the timing information over all (~20) detected photons shrinks the timing jitter to ~11 ps. To verify these reconstruction results, we applied a TOP-position calibration, where the average LOP collected by each sensor pixel and the average velocity of Cherenkov photons are used to calculate the TOP, and the same results are obtained from simulation truth with no further correction (such as the dispersion effect). It is noted that in the TOF distribution plot, a low (but visible) long tail shows on both sides of the main peak. The tail is mainly caused by secondary particles along with the primary pion, mostly δ-electrons.

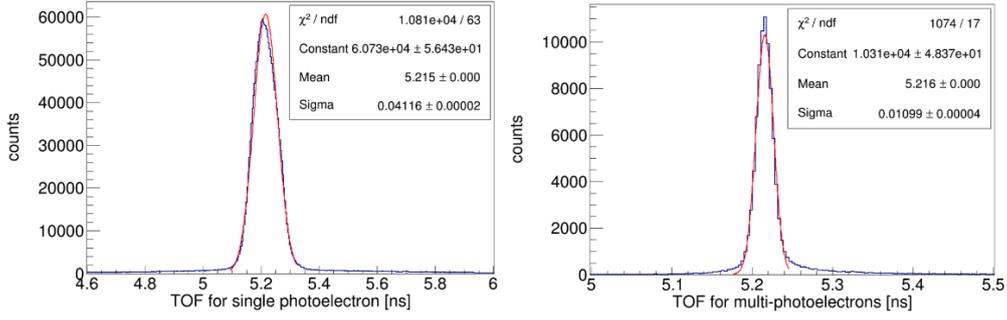

**Figure 9.** Intrinsic TOF resolution from DTOF reconstruction for the single photoelectron and combining all photoelectrons.

### 5.3 PID capability

Given a particle-species hypothesis, the TOF measurement can be calculated as
$$TOF_h = T - TOP_h - T_0,$$
where *h* denotes the different hypothesis particles. Then we can find a deviation between these TOF measurement with different particle hypotheses, which is mainly owing to the TOP reconstruction deviation, shown as:
$$TOF_{h1} - TOF_{h2} = TOP_{h2} - TOP_{h1}$$

The expectations of each hypothesis particle are subsequently compared to study the PID capability of the DTOF. Fig. 10 shows the reconstructed TOF distributions of both pions and kaons at 2 GeV/c for the SPE and combining all photoelectrons when convoluting all the contributing factors, including the TTS of 70 ps and the T0 uncertainty of 40 ps. One can easily find if the particle hypothesis is correct, the reconstructed TOF peak is at its right position, with a resolution of ~50 ps, as shown in Fig. 10 (right). However, if the hypothesis is untrue, the reconstructed TOF peak can be shifted with respect to its expectation. This shift further enlarges the separation between pion and kaon TOF peaks, which may have a positive impact on PID power. By directly comparing the TOF information, a ~3σ separation power for π/K at 2 GeV/c is achieved, fulfilling the required PID capability of the DTOF. Furthermore, the separation power becomes stronger when we compare the reconstructed TOFs of various hypotheses for the same set of particles. For either the pion or kaon samples at 2 GeV/c, the separation power reaches ~4σ by comparing the TOF distributions under π or K hypotheses.

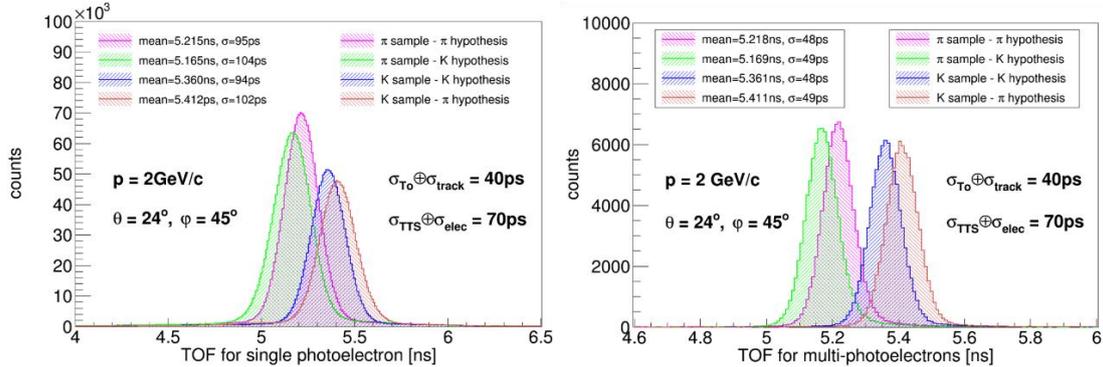

**Figure 10.** TOF PID capability of the DTOF detector for the π/K separation at 2 GeV/c, for the single photo-electron (left) and combining all photoelectrons (right).

To further evaluate the PID capability of the DTOF, we applied the likelihood method. The likelihood function can be expressed as

$$\mathcal{L}_h = \prod_{i=1}^{n} f_h(TOF_i^h),$$

where $h$ denotes the hadron species (in our case, $\pi$ and K), and $i$ accounts for each detected photon. The probability density function $f_h$ is taken as the Gaussian fit to the expected TOF distribution (as in Fig. 10 (left)) after normalization, in addition to a constant background of 0.05. The reconstructed $Log\,\mathcal{L}_\pi - Log\mathcal{L}_K$ for 2 GeV/c $\pi$ and K emitted at different angles shown in Fig. 11. Despite the very different particle directions, the separation power of the DTOF is similar ~4σ or better in all the DTOF sensitive areas, as shown in Fig. 12 (left). The π/K separation power at different momenta is also shown in Fig. 12, indicating a better performance of the DTOF at a low momentum. Notably, even with a TTS of 100 ps, the π/K resolution remains at 3.89σ (p = 2 GeV/c, θ = 24°, and φ = 45°). This still meets the requirement of 3σ π/K separation power.

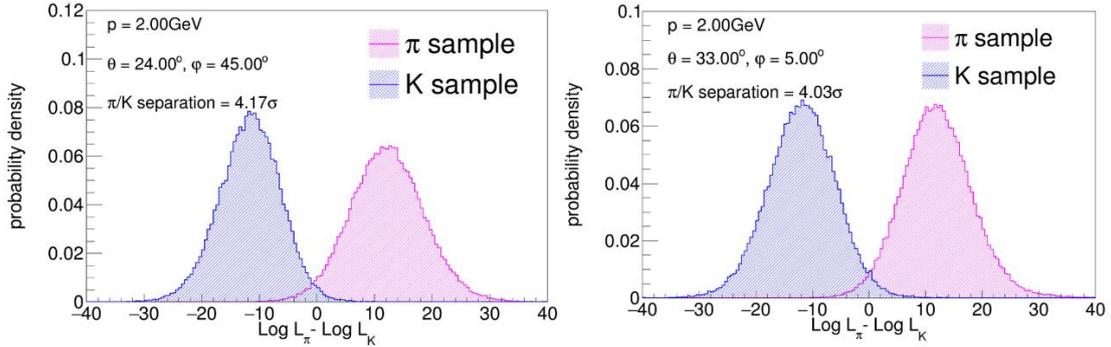

**Figure 11.** Likelihood PID capability of the DTOF detector for the π/K separation at 2 GeV/c, emitted at different angles.

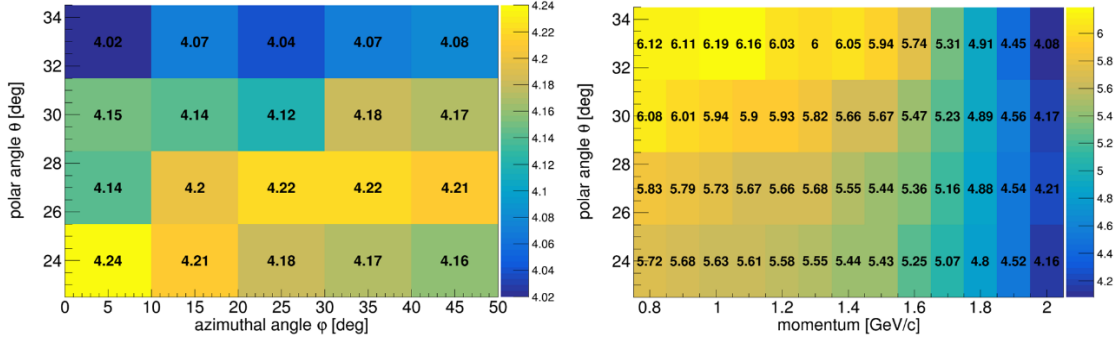

**Figure 12.** 12 π/K separation power: p = 2 GeV/c at different directions (left) and φ = 45° at different momenta (right).

## 6. Background impact

In experiments involving high luminosity machines such as the STCF, the intensive background has significant impact on the performance of the various detectors used. The background particles obtained from a specific machine-detector interface (MDI) [23] are used as inputs for the Geant4 simulation of the DTOF to estimate the effect of background on the DTOF performance. The background particles hitting the DTOF are mainly gamma photons, electrons, and a few hadrons, and the hit rate of all the particles on the DTOF is approximately $7 \times 10^9$ Hz. The energies of most backgrounds are smaller than the Cherenkov threshold. Only a part of the charge particles or secondary particles produced in the DTOF radiator generates the Cherenkov light. These backgrounds include two main parts: the beam-induced background (~75%) and the physical background (~25%). The beam-induced background is uniformly distributed in time and unrelated to the beam bunch crossing, whereas the physical background is related to the collisions during bunch crossing (once per 8 ns) and exhibits a characteristic time structure. The overall time distribution of background hit on the DTOF can be obtained by combining two kinds of background time structures as shown in Fig. 13.

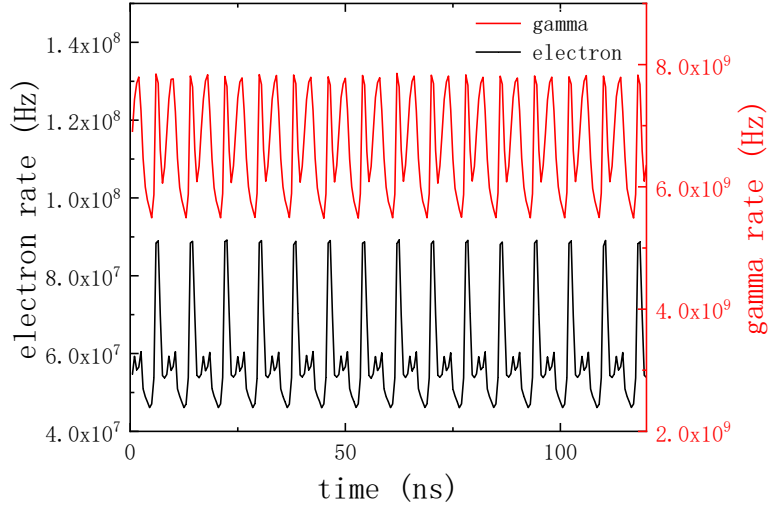

**Figure 13.** Overall time distribution of background electrons and gamma photons on the DTOF.

In the Geant4 simulation, the hit position, direction, and energy of the background particles are extracted from the background samples. The time window of the signal acquisition is 100 ns, which is within the interval of [-40 ns, 60 ns], thereby positioning the real signal in the middle of the time window. In this time window, the number of background particles is derived using Poisson distribution. The hit time is sampled from a uniform distribution rather than the specific distribution as shown in Fig. 13. This is because the different time distribution has little effect on the results, and a uniformity distribution can significantly simplify the sampling process.

Pions and kaons are generated in the Geant4 simulation along with the background samples to study the effect of the background. It was found that the background may lead to an increase in the number of photoelectrons detected by the DTOF in a single event, resulting in an increased possibility of multiple hits in a single channel. The multiple hits are corrected during data processing. In other words, in the time window of [-40 ns, 60 ns], when a single channel contains multiple hits, only the earliest photoelectron signal is taken and all the other hits are disregarded. The average number of photoelectrons for a pion at p = 2 GeV/c is approximately 33 when taking the background hits (approximately 11 in 100 ns window) into account and applying the correction of multiple hits in a single channel. Notably, if the average number of background photoelectrons is 11 for a time window of 100 ns, with an assumed MCP-PMT gain of $10^6$, the average accumulated charge density on the MCP-PMT anode is approximately 11 $C/cm^2$ over 10-year STCF operation (50% run time). This indicates that the aging of the photocathode, i.e. the loss of quantum efficiency owing to ion backflow, poses a challenge to the lifetime of the MCP-PMT [21,22]. However, a new MDI with more shielding can further suppress the background and extend the lifetime of MCP-PMT in the future.

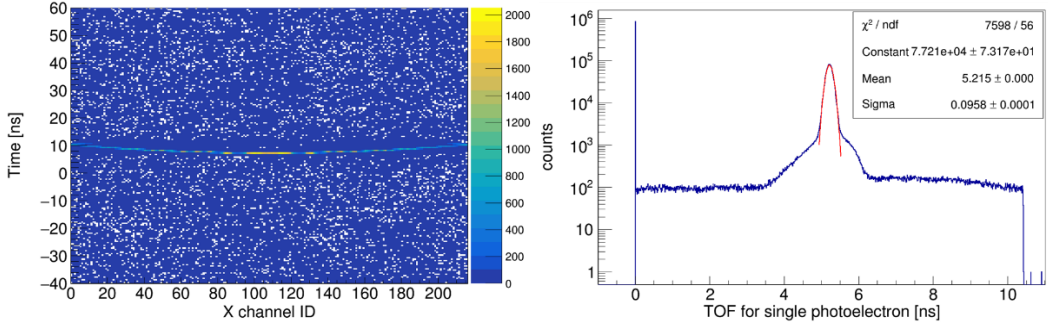

**Figure 14.** 2-D time-position map of the DTOF (left) and the TOF distribution of the single photo-electron signal (right), both with background hits.

Fig. 14 (left) depicts the 2-D time-position map of the DTOF hits by pions at p = 2 GeV/c, θ = 24° and φ = 45°. It can be seen that the hits by the background particles are uniformly distributed throughout the phase space, whereas the real signal hits are concentrated in the form of a band. After time reconstruction, the TOF distribution of the SPE signal can be obtained, as shown in Fig. 14 (right). The reconstructed TOF of the real signal in the figure shows a Gaussian distribution (mean 5.22 ns, sigma ~96 ps, with the convolution of all contributing factors), whereas the TOF of the background particles are distributed rather uniformly. Some SPEs do not meet the reconstruction conditions and are taken as background during the reconstruction process. These are represented as the peak with zero TOF in the figure. Owing to the uniform distribution of the reconstructed background signal, the influence of the background can be largely eliminated by using the maximum likelihood method. The π/K resolution is found to be 4.15σ, as shown in Fig. 15. Considering the Poisson fluctuation of background count, the π/K resolution remains at 4.12σ even under an extreme condition of three standard deviations above the average background level. This PID performance, however, meets the requirement of 3σ π/K separation for the DTOF, i.e. the background effect on π/K identification is fairly small.

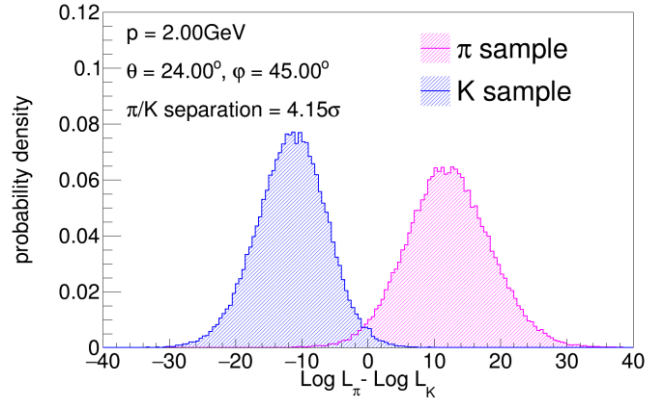

**Figure 15.** π/K identification capability (at p = 2 GeV/c) by DTOF with background hits.

## 7. Conclusion

A conceptual design of the DTOF detector is proposed as an endcap PID detector for the experiment at STCF to provide an effective π/K/p identification. The main sources contributing to the timing uncertainty of the DTOF detector are analyzed. Our study shows that the timing uncertainty of the photon sensor largely contributes to the intrinsic time resolution for short photon propagation length, whereas the dispersion effect gradually becomes the dominant factor when the transmission distance of the photon is large. An optimum quadrantal radiator with a thickness of 15 mm and an absorber attached to its outer side surface, is chosen as our baseline design. The performance of the DTOF is investigated through a Geant4 simulation, and a reconstruction algorithm of the DTOF is developed. The Geant4 simulation indicates an overall reconstructed TOF time resolution of ~50 ps with an average of ~20 photons detected by the MCP-PMT arrays when all the contributing factors are converted. It is worth noting that the uncertainty of $T_0$ dominates the overall timing error; therefore, an optimal design of the STCF bunch size is crucial. By applying the likelihood method, a π/K separation power of DTOF of ~4σ or better at a momentum of 2 GeV/c is achieved over the entire DTOF sensitive area. This fulfils the requirement for the PID detector at the STCF. Furthermore, the background effect on the π/K identification of the DTOF detector is fairly small, and the average accumulated charge density on the MCP-PMT anode is approximately 11 C/cm$^2$ over a 10-year STCF operation. This poses a challenge to the lifetime of the MCP-PMT.

## Acknowledgments

This work was supported by the Chinese Academy of Sciences [Grant NO.: GJJSTD20200008]; the National Natural Science Foundation of China [Grant No. U1932202, 11775217]; and the Double First-Class university project foundation of USTC. The authors thank Hefei Comprehensive National Science Center for their strong support.